# Choking non-local magnetic damping in exchange biased ferromagnets

Takahiro Moriyama[a)], Kent Oda, and Teruo Ono

*Institute for Chemical Research, Kyoto University, Gokasho, Uji, Kyoto, 611-0011, Japan.*

[a)] corresponding to :*mtaka@scl.kyoto-u.ac.jp*

**Abstract**

We investigated the temperature dependence of the magnetic damping in the exchange biased Pt/ $Fe_{50}Mn_{50}$ /$Fe_{20}Ni_{80}$ /$SiO_x$ multilayers. In samples having a strong exchange bias, we observed a drastic decrease of the magnetic damping of the FeNi with increasing temperature up to the blocking temperature. The results essentially indicate that the non-local enhancement of the magnetic damping can be choked by the adjacent antiferromagnet and its temperature dependent exchange bias. We also pointed out that such a strong temperature dependent damping may be very beneficial for spintronic applications.

The Gilbert damping constant, $\alpha$, appearing in the Landau-Lifshitz-Gilbert equation is one of the important parameters characterizing the magnetization dynamics [1]. It influences the switching speed of the magnetization [2, 3] and also determines the threshold current for various spin-torque-related phenomena [4, 5, 6, 7, 8]. With those numerous examples, it is probably not an exaggeration to say that the performance of every spintronic device relies on $\alpha$. It is, however, unfortunate that the magnetic damping is one of the least controllable magnetic parameters among those including magnetic anisotropy, saturation magnetization, and gyromagnetic ratio which a material engineering as well as some exogenous controls (e.g. temperature, electric field bias, structural strain, etc.) [9, 10, 11, 12, 13, 14, 15] can easily tune over a wide range. Although some efforts have been made to minimize $\alpha$ by carefully engineering the band structure [16], an exogenous control of it, which is very much desirable for spintronic applications, has rarely been explored.

The magnetic damping is rooted in various relaxation processes of the spin angular momentum [17]. The majority of such processes locally occur within the magnetic material via, *e.g.*, spin-phonon and spin-spin relaxations [18]. On the other hand, an interaction between the magnetization dynamics and the itinerant electron spin, *i.e.* the spin pumping effect [19], results in a non-local magnetic damping of the spin angular momentum. In other words, when the spin angular momentum carried by the itinerant electron diffuses into an adjacent non-magnetic material, an additional damping enhancement can occur depending on the spin relaxation properties of the non-magnetic material. For instance, a strong damping enhancement due to the spin pumping effect has been observed Pt/ ferromagnet (FM) multilayers in which the Pt works as a good spin dissipative material [20]. This non-local damping enhancement has been recently revisited for the system of the exchange biased FM/antiferromagnet (AFM) multialyers [21]. In Ref. 21, it was shown that the non-local damping was modified by the Néel order in the AFM. In particular, the linear relation between the strength of the exchange bias (EB) and the enhanced damping strongly indicates that an additional spin relaxation takes effect due to the orientation of the Néel order. Furthermore, the relatively gradual AFM thickness dependence of the damping enhancement, suggesting a spin current transmission through the AFM [22, 23, 24, 25, 26, 27, 28], was also certainly intriguing.

From another point of view, these results indeed offer a control of the magnetic damping by the magnetic order in the adjacent AFM. It is therefore worthwhile to further investigate and understand the magnetic damping in exchange biased FM/AFM systems and to gain a control of it for the sake of advancement in spintronic applications. In this work, we extended our investigations to the temperature dependence of the non-local damping in the exchange biased AFM/FM multilayer. We particularly paid attention to the damping variations toward the blocking temperature $T_B$ of the EB. It is found that the non-local magnetic damping drastically decreases with increasing temperature when EB is strong. Moreover, the linear relation between the strength of the EB and the enhanced damping, previously found in terms of the AFM thickness dependence of EB [21], is found to also hold in terms of the temperature dependence.

We investigated Pt 5 nm/ $Fe_{50}Mn_{50}$ $t_{FeMn}$ nm/$Fe_{20}Ni_{80}$ 4 nm/$SiO_x$ 2nm ($t_{FeMn}$ = 0, 3, 20, and 60 nm) multilayers grown on a thermally oxidized Si substrate as illustrated in Fig. 1 (a). The samples were photolithographically patterned into a 10 μm wide strip attached to a coplanar waveguide made of Ti/Au layer facilitating a high frequency measurement. EB was set by a field cooling process with an annealing temperature of $T_{FC}$

= 200 C° and a field of 2 kOe. No irreversible degradations of the film, *e.g.* due to atom diffusions, were found in this field cooling process [21]. High sensitivity ferromagnetic resonance (FMR) measurements were carried out by a homodyne detection technique [29, 30, 31] with an identical circuitry and methodology used in Ref. 21. The homodyne voltage $V_{dc}$ were measured with a fixed frequency $f$ of the r.f. current and with a swept external magnetic field applied either along or against the direction of the exchange bias field $H_{eb}$ which are denoted by "along EB" and "against EB", respectively. The FMR measurements were at first performed at 298K and the measurement temperature was stepped up to 393 K.

Figure 1 (b) shows representative FMR spectra obtained at 298 K and 393 K for $t_{FeMn}$ = 20 nm with $f$ = 8 GHz and the external field along and against EB. The displacement between the two spectra at 298 K with the fields along EB and against EB manifests a unidirectional magnetic anisotropy, i.e. the exchange bias field. One can also see that the exchange bias is diminished at 393 K. FMR spectra are fitted by the combination of symmetric and anti-symmetric Lorentzians by which the resonant parameters, such as the resonant field $H_{res}$ and the spectral linewidth $\varDelta H$, are extracted [29].

Figure 2 shows $f$ vs. $H_{res}$ and $\varDelta H$ vs. $f$ at 298 K and 393 K for $t_{FeMn}$ = 0, 3, 20, and 60 nm. In order to extract the effective demagnetizing field $4\pi M_{\text{eff}}$ and the exchange bias field $H_{eb}$, the $f$ vs. $H_{res}$ curves are fitted by the Kittel's equation $f = (\gamma/2\pi)\sqrt{(H_{\text{res}} + H_u)(H_{\text{res}} + 4\pi M_{\text{eff}})}$, where $\gamma$ is the gyromagnetic ratio and $H_u$ is the anisotropy field from which both uniaxial anisotropy field and the exchange bias field $H_{eb}$ are derived. The Gilbert damping $\alpha$ is extracted from the $\varDelta H$ vs. $f$ plot by using $\Delta H = \Delta H_0 + 2\pi\alpha f/\gamma$ [32], where $\Delta H_0$ is a frequency independent-linewidth known as the inhomogeneous broadening [33]. We note that our linewidth analyses explicitly separate the frequency dependent damping, or the Gilbert damping $\alpha$, from the independent one, or inhomogeneous broadening which may be related to the two-magnon scattering at the FM/AFM interface [34,35]. As shown in Fig. 2, all the samples exhibit a good Kittel's fitting as well as a linear fitting for the $\alpha$ extraction. We note that, for $t_{FeMn}$ = 3 nm at 298 K, there was an exceptionally large uniaxial magnetic anisotropy field of 390 Oe, which may lead to significantly large $\Delta H_0$ compared to samples with other $t_{FeMn}$ which essentially read $\Delta H_0$ = 0. All the discussions about the magnetic damping thereafter will be referring to $\alpha$ which reflects the intrinsic property of the system [33].

Figures 3 (a), (b), and (c) summarize $4\pi M_{\text{eff}}$, $H_{eb}$, and $\alpha$, respectively, as a function of temperature. For all the samples, $4\pi M_{\text{eff}}$ is ranging from 0.88 to 0.96 Tesla which is considered reasonable for the $Fe_{20}Ni_{80}$ with a possible interfacial perpendicular anisotropy [36]. $4\pi M_{\text{eff}}$ overall decreases by ~10 % with increasing temperature from 298 to 393 K, which can be attributed to the thermal decay of the magnetization as well as a change in the interfacial anisotropy. The exchange bias field is found to be largest $H_{eb}$ = 79 Oe with $t_{FeMn}$ = 20 nm and it monotonically decreases with increasing temperature. The blocking temperature for the exchange bias is found to be $T_B$ = 393 K and 373 K for $t_{FeMn}$ = 20 and 60 nm, respectively. No exchange bias field was observed for $t_{FeMn}$ = 0 and 3 nm in the measurement temperature range. It is remarkable that, for different $t_{FeMn}$, $\alpha$ behaves quite differently with respect to temperature. Namely, the samples with $t_{FeMn}$ = 0 and 60 nm show a slight upward trend of $\alpha$ with increasing temperature but one with $t_{FeMn}$ = 3 shows almost constant with respect to temperature. It is noticeable that $\alpha$ for $t_{FeMn}$ = 20 nm undergoes a drastic decrease, almost by a half, with increasing temperature up to $T_B$. We also emphasize that $\alpha$ measured with the field along

EB was found to be always smaller than that measured with the field against EB.

Figure 3 (d) plots $\Delta\alpha$ as a function of $H_{eb}$ where $\Delta\alpha$ is the difference of $\alpha$ measured with the field along EB and against EB. It is found that $\Delta\alpha$ increases lineally with respect to $H_{eb}$. Although $H_{eb}$ varies due to temperature in this present case, the observation of the linear relation between $H_{eb}$ and $\Delta\alpha$ is essentially same as what was observed previously with respect to $H_{eb}$ which varies with respect to $t_{FeMn}$ [21]. The inset of Fig. 3 (d) shows $\alpha$ as a function of the field angle $\varphi$ with respect to the exchange bias direction ($\varphi = 0$ corresponds to the direction along EB.) measured for $t_{FeMn} = 20$ nm at 298 and 343 K, which depicts the $\varphi$ dependence of $\alpha$ similar to that observed in the previous report [21].

Now, one finds two peculiar points in our results, especially when focusing on $t_{FeMn} = 20$ nm. One is the strong temperature dependence of $\alpha$ with the exchange bias. The other is the temperature dependence of $\Delta\alpha$ which results in the linear relation between $\Delta\alpha$ and $H_{eb}$. While the latter is consistent with our previous report and originates from the Néel order twisting [21], the former is quite intriguing in both physics and application points of views. Looking at Figs. 3 (c) and (d), one can presume that the reduction of $\alpha$ may be correlated with the strength of the exchange bias.

It has been shown that the intrinsic damping of a single layer of FeNi only slightly varies in this temperature range [37]. A non-local damping enhancement by an adjacent Pt due to the spin pumping effect should not significantly vary since the relevant parameters such as the mixing conductance and the spin diffusion length in Pt do not vary much in this temperature range. Very little temperature variation of $\alpha$ in our control sample ($t_{FeMn} = 0$ nm) is therefore consistent with those previous observations as well as the expectations. It is now very clear that the observed temperature dependence of $\alpha$ with $t_{FeMn}$ =20 nm is solely associated with the FeMn insertion and the exchange bias between the FeMn and FeNi [38] (Also, see Supplementary Information).

The mechanism of the exchange bias is generally explained by two major factors [39]; *i.e.* the exchange coupling at the FM/AFM interface and the magnetic anisotropy energy of the AFM. Because a thermal agitation effectively reduces the magnetic anisotropy, the exchange bias field diminishes toward $T_B$ at which the effective magnetic anisotropy essentially becomes zero. Considering the spin transmission in antiferromagnets by the form of the Néel order dynamics, there is a negative correlation between the magnetic anisotropy and the spin diffusion length (or the so-called healing length for the spin transmission) $\lambda_{FeMn}$ [24, 40]. Therefore, our observations shown in Figs. 3(b) and (c) can be considered representing a negative correlation between $\alpha$ and $\lambda_{FeMn}$.

According to Ref. 40, $\alpha$ as a function of $\lambda_{FeMn}$ behaves differently depending on either the strong or weak damping limit of the antiferromagnet. Our experimental results are more consistent with the strong damping limit in which $\alpha$ increases with decreasing $\lambda_{FeMn}$. We should note here that the present analysis neglects a direct thermal effect on the Néel order dynamics, such as thermal magnons, which could be more important than considering the effective magnetic anisotropy but is difficult to be taken into account analytically at this moment.

Finally, we highlight our results in the engineering point of view. As we pointed out above, the magnetic damping of FeNi should generally exhibit only a slight temperature dependence, at around room temperature, regardless of the local or non-local mechanisms. Our results suggest that, by making use of the exchange biased bilayer, one can drastically "choke" the non-local enhancement of the magnetic damping in a relatively narrow temperature range. For those spintronic applications involving a Joule heating upon the operation (*e.g.* the spin torque magnetic memory [6] requires quite a bit of current density to write information bits which heats up the ferromagnetic bits.), this temperature control of $\alpha$ may be beneficial since it is reduced only when they are in operation.

In summary, we investigated the temperature dependence of the magnetic damping in the exchange biased Pt/ $Fe_{50}Mn_{50}$ /$Fe_{20}Ni_{80}$ /$SiO_x$ multilayer in the temperature range of 298 ~ 393 K. We reconfirmed the linear relation between $H_{eb}$ and $\Delta\alpha$, which have been seen in our previous report [21], by the present experimental approach which is different from that in the previous report. In samples having a strong exchange bias, we observed the drastic decrease of the magnetic damping with increasing temperature up to the blocking temperature. The results lead us to the negative correlation between the magnetic anisotropy of the FeMn and the damping enhancement, implying the mechanism of spin current transmission through the FeMn with a strong damping limit [40]. Furthermore, we pointed out that this strong temperature dependent damping may be very beneficial for spintronic applications.

Acknowledgements

This work was supported in part by JSPS KAKENHI Grant Numbers 26870300, 17H04924, 15H05702, and 17H05181 ("Nano Spin Conversion Science"). We also acknowledge the support from Center for Spintronics Research Network (CSRN).

**Figure captions:**

Fig. 1 (a) Schematic illustration of the layer structure under investigation by the homodyne FMR measurement. (b) Representative FMR spectra (the homodyne voltage $V_{dc}$ as a function of the applied field either along (red markers) or against EB (blue markers)) for $t_{FeMn}$ = 20 nm at $f$ = 8 GHz measured at 298 K (upper panel) and 393 K (lower panel). The continuous lines are the fitting by the combination of symmetric and anti-symmetric Lorentzians.

Fig. 2 Resonant field $H_{res}$ and the spectral linewidth $\Delta H$ as a function of frequency $f$ measured at 298 K and 393 K for (a, e) $t_{FeMn}$ = 0 nm, (b, f) $t_{FeMn}$ = 3 nm, (c, g) $t_{FeMn}$ = 20 nm, and (d, h) $t_{FeMn}$ = 60 nm. The data with the applied field along and against EB are plotted. The continuous lines are the fitting by the Kittel's equation and $\Delta H = \Delta H_0 + 2\pi\alpha f/\gamma$ described in the main text.

Fig. 3 (a) $4\pi M_{eff}$ as a function of temperature, (b) $H_{eb}$ as a function of temperature, (c) $\alpha$ as a function of temperature, and (d) $\Delta\alpha$ as a function of $H_{eb}$ for $t_{FeMn}$ = 0 nm (dark blue), 3 nm, (light blue), 20 nm (green), and 60 nm (red). In panel (c), $\alpha$ obtained with the applied field along EB (upward triangle) and against EB (downward triangle) are separately shown. The inset of (d) shows $\alpha$ as a function of the field angle $\varphi$ with respect to the exchange bias direction for $t_{FeMn}$ = 20 nm. (note that $\varphi$ = 0° and 180° correspond to "along EB" and "against EB", respectively.).

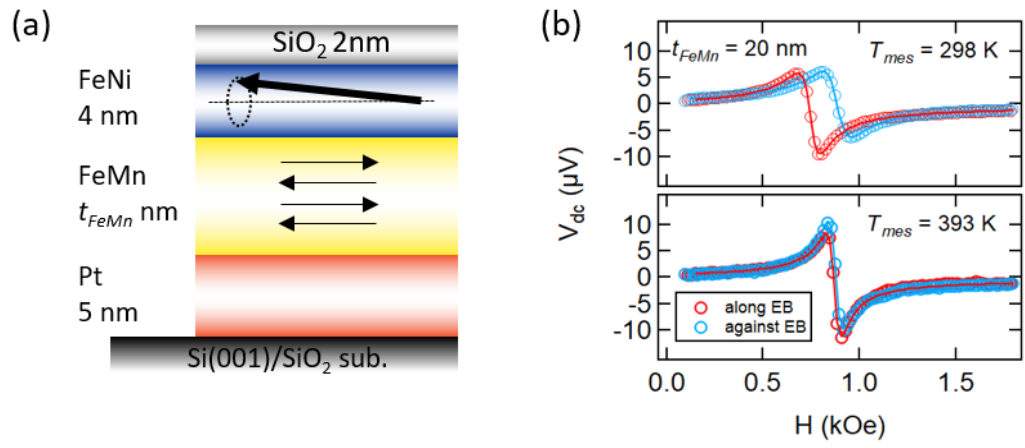

Figure 1 Moriyama et al.

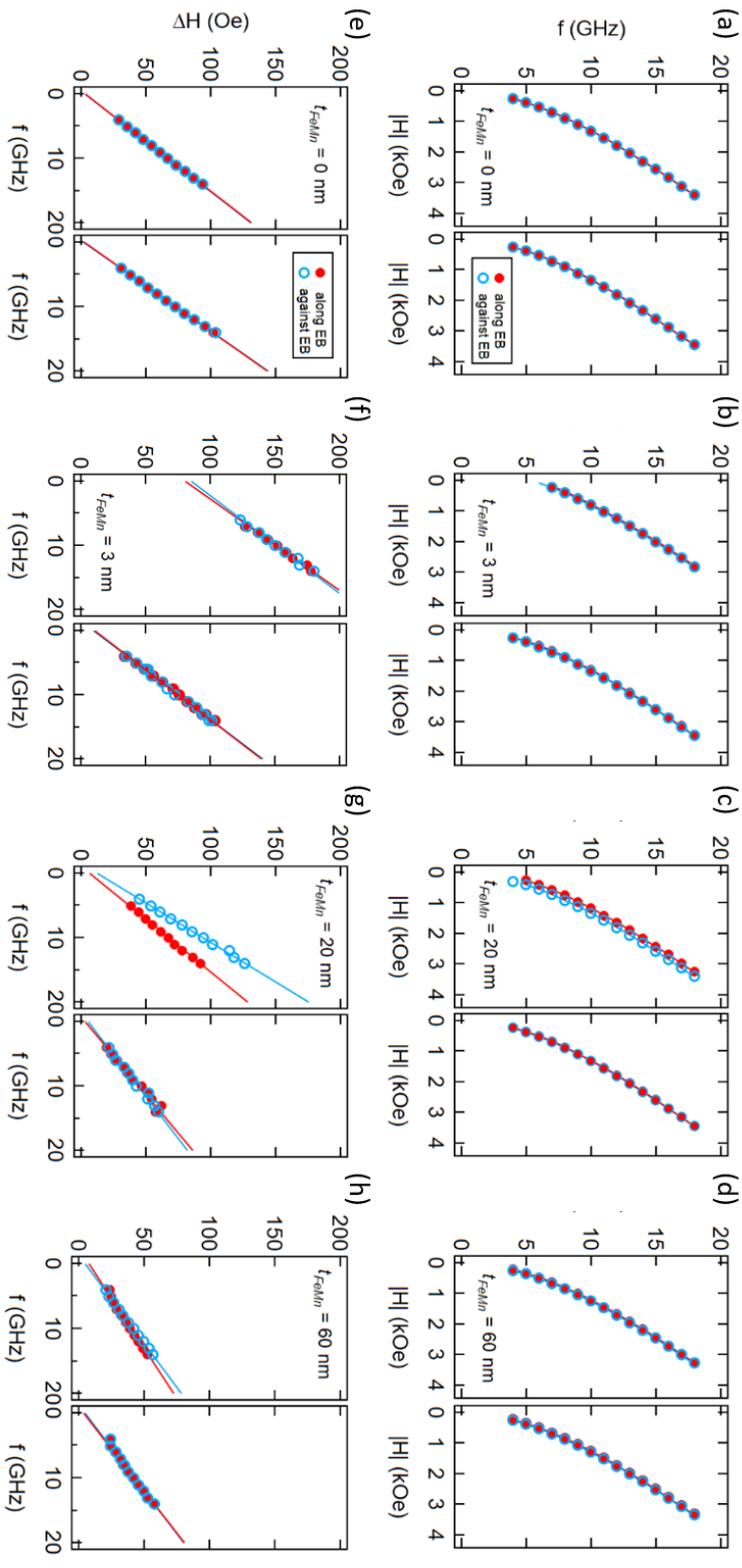

Figure 2 Moriyama et al.

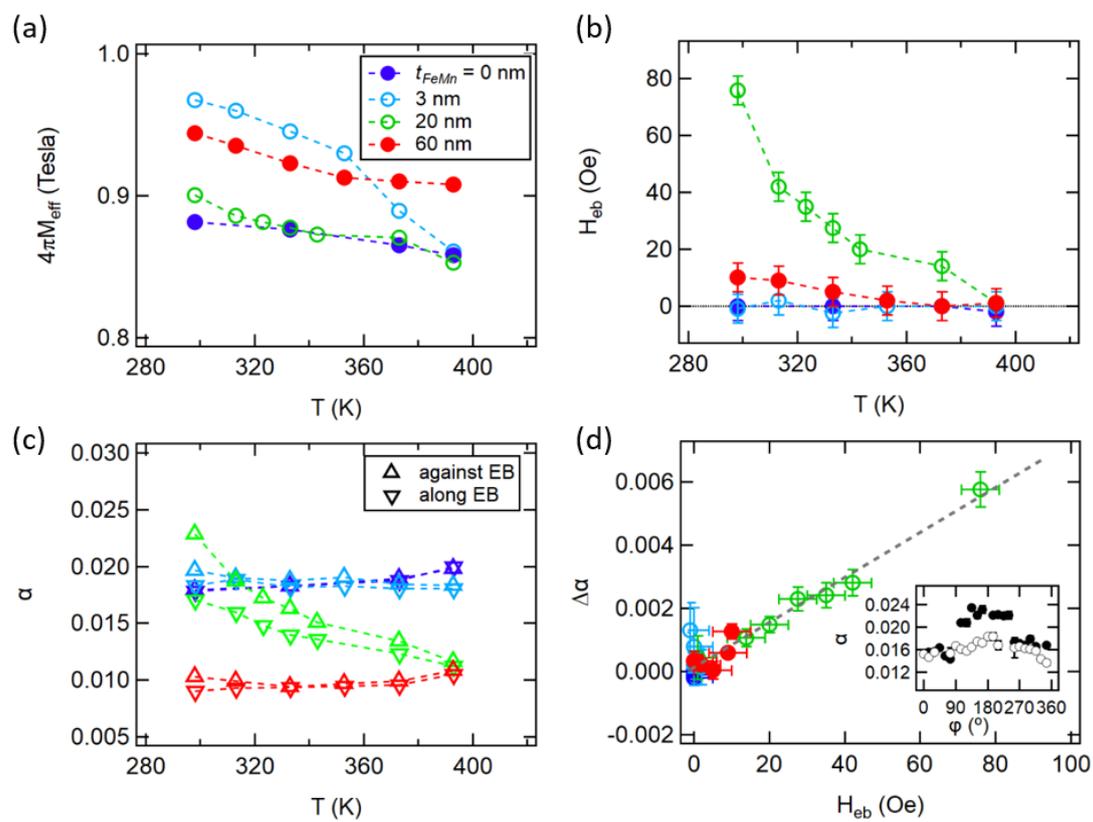

Figure 3 Moriyama et al.

**References:**

[1] T. L. Gilbert, *Phys. Rev.* **100**, 1243 (1955).
[2] C.H. Back et al., *Phys. Rev. Lett.* **81**, 3251 (1998).
[3] M. Bauer, R. Lopusnik, J. Fassbender, and B. Hillebrands, *J. Magn. Magn. Mater.* **218**, 165 (2000).
[4] D. C. Ralph and M. D. Stiles, *J. Magn. Magn. Mater.* **320**, 1190 (2008).
[5] J. Z. Sun, *Phys. Rev. B* **62**, 570 (2000).
[6] J. Z. Sun and D. C. Ralph, *J. Magn. Magn. Mater.* **320**, 1227 (2008).
[7] S. I. Klselev, J. C. Sankey, I. N. Krivorotov, N. C. Emley, R. J. Schoelkopf, R. A. Buhrman, and D. C. Ralph, *Nature* **425**, 380 (2003).
[8] Z. Li and S. Zhang, *Phys. Rev. B* **70**, 1 (2004).
[9] G. H. O. Daalderop, P. J. Kelly, and F. J. A. Den Broeder, *Phys. Rev. Lett.* **68**, 682 (1992).
[10] C. D. Stanciu, A. V. Kimel, F. Hansteen, A. Tsukamoto, A. Itoh, A. Kirilyuk, and Th. Rasing, *Phys. Rev. B* **73**, 220402 (2006)
[11] M. Weisheit, M. Weisheit, S. Fähler, A. Marty, Y. Souche, C. Poinsignon, and D. Givord, *Science* **349**, (2011).
[12] T. Maruyama, Y. Shiota, T. Nozaki, K. Ohta, N. Toda, M. Mizuguchi, A. A. Tulapurkar, T. Shinjo, M. Shiraishi, S. Mizukami, Y. Ando, and Y. Suzuki, *Nat. Nanotechnol.* **4**, 158 (2009).
[13] D. Chiba, S. Fukami, K. Shimamura, N. Ishiwata, K. Kobayashi, and T. Ono, *Nat. Mater.* **10**, 853 (2011).
[14] K.-J. Kim, S. K. Kim, T. Tono, S.-H. Oh, T. Okuno, W. S. Ham, Y. Hirata, S. Kim, G. Go, Y. Tserkovnyak, A. Tsukamoto, T. Moriyama, K.-J. Lee, and T. Ono, *Nat. Mater.* **16**, 1187 (2017).
[15] S. Ota, A. Ando, and D. Chiba, *Nat. Elec.* **1**, 124 (2018).
[16] M. A. W. Schoen, D. Thonig, M. L. Schneider, T. J. Silva, H. T. Nembach, O. Eriksson, O. Karis, and J. M. Shaw, *Nat. Phys.* **12**, 839 (2016).
[17] Magnetization Oscillations and Waves by A.G. Gurevich and G.A. Melkov, 1996 by CRC Press.
[18] Anderson, *Phys. Rev.* **88**, 1214 (1952).
[19] Y. Tserkovnyak, A. Brataas, G. E. W. Bauer, and B. I. Halperin, *Rev. Mod. Phys.* **77**, 1375 (2005).
[20] S. Mizukami, Y. Ando, and T. Miyazaki, *Phys. Rev. B* **66**, 104413 (2002).
[21] T. Moriyama, M. Kamiya, K. Oda, K. Tanaka, K-J. Kim, and T. Ono, *Phys. Rev. Lett.* **119,** 267204 (2017).
[22] H. Wang, C. Du, P. C. Hammel, and F. Yang, *Phys. Rev. Lett.* **113,** 097202 (2014).
[23] C. Hahn, G. Loubens, V. V. Naletov, J. B. Youssef, O. Klein, and M. Viret, *Europhys. Lett.* **108,** 57005 (2014).
[24] T. Moriyama, S. Takei, M. Nagata, Y. Yoshimura, N. Matsuzaki, T. Terashima, Y. Tserkovnyak, and T. Ono, *Appl. Phys. Lett.* **106,** 162406 (2015).
[25] L. Frangou, S. Oyarzún, S. Auffret, L. Vila, S. Gambarelli, and V. Baltz, *Phys. Rev. Lett.* **116**, 077203 (2016).
[26] Z. Qiu, J. Li, D. Hou, E. Arenholz, A. T. N'Diaye, A. Tan, K.-i. Uchida, K. Sato, S. Okamoto, Y. Tserkovnyak, Z. Q. Qiu, and E. Saitoh, *Nat. Commun.* **7**, 12670 (2016).
[27] W. Lin, K. Chen, S. Zhang, and C. L. Chien, *Phys. Rev. Lett.* **116**, 186601 (2016).
[28] T Ikebuchi, T Moriyama, H Mizuno, K Oda, and T Ono, *Appl. Phys. Exp.* **11**, 073003 (2018).
[29] L. Liu, T. Moriyama, D. C. Ralph, and R. A. Buhrman, *Phys. Rev. Lett.* **106,** 036601 (2011).
[30] H. Mizuno, T. Moriyama, M. Kawaguchi, M. Nagata, K. Tanaka, T. Koyama, D. Chiba, and

T. Ono, *Appl. Phys. Exp.* **8,** 073003 (2015).
[31] T. Ikebuchi, T. Moriyama, Y. Shiota, and T. Ono, *Appl. Phys. Exp.* **11,** 053008 (2018).
[32] B. Heinrich, J. F. Cochran, and R. Hasegawa, *J. Appl. Phys.* **57,** 3690 (1985).
[33] R. Arias and D. L. Mills, *Phys. Rev. B* **60**, 7395 (1999).
[34] S. M. Rezende, A. Azevedo, M. A. Lucena, and F. M. de Aguiar, *Phys. Rev. B* **63**, 214418 (2001).
[35] M. C. Weber, H. Nembach, and B. Hillebrands, J. Fassbender, *J. of Appl. Phys.* **97**, 10A701 (2005).
[36] J.-M. Beaujour, D. Ravelosona, I. Tudosa, E. E. Fullerton, and A. D. Kent, *Phys. Rev. B* **80**, 180415 (2009).
[37] Y. Zhao, Q. Song, S. H. Yang, T. Su, W. Yuan, S. S. P. Parkin, J. Shi, and W. Han, *Sci. Rep.* **6**, 4 (2016).
[38] We also verified the Pt underlayer does not play a role in this regard. See Supplementary Information for extended investigations [URL].
[39] D. Mauri, H.C. Siegmann, P. S. Bagus, and E. Kay, *J. Appl. Phys.* **62**, 3047 (1987).
[40] S. Takei, T. Moriyama, T. Ono, and Y. Tserkovnyak, *Phys. Rev. B* **92**, 020409(R) (2015).